\documentstyle[aps,graphicx,amssymb,eqsecnum,epsfig,amsmath]{revtex}
\title{Planform selection in two-layer Benard-Marangoni convection}

\author{A. Engel\thanks{email:andreas.engel@physik.uni-magdeburg.de} $^a$ 
        and J. B. Swift$^b$\\[2ex]
      $^a$ {\small Institut f\"ur Theoretische Physik,
        Otto-von-Guericke Universit\"at, Postfach 4120, D-39016
        Magdeburg, Germany}\\
      $^b$ {\small Center for Nonlinear Dynamics, University of
        Texas at Austin, Austin, TX 78712, USA}}

\newcommand{\te}{\theta}
\newcommand{\Te}{\Theta}
\newcommand{\la}{\lambda}
\newcommand{\La}{\Lambda}
\newcommand{\ka}{\kappa}
\newcommand{\ve}{\boldsymbol{v}}
\newcommand{\Ve}{\boldsymbol{V}}
\newcommand{\pa}{\partial}
\newcommand{\na}{\nabla}
\newcommand{\ez}{\boldsymbol{e}_z}
\newcommand{\eps}{\varepsilon}
\newcommand{\ph}{\varphi}
\newcommand{\bk}{\boldsymbol{k}}
\newcommand{\br}{\boldsymbol{r}}
\newcommand{\bq}{\boldsymbol{q}}
\newcommand{\bp}{\boldsymbol{p}}
\newcommand{\para}{\text{par}}

\begin{document}    

\maketitle

\begin{abstract}
B\'enard-Marangoni convection in a system of two superimposed liquids is
investigated theoretically. Extending previous studies the complete
hydrodynamics of both layers is treated and buoyancy is consistently taken
into account. The planform selection problem between rolls, squares and
hexagons is investigated by explicitly calculating the coefficients of an
appropriate amplitude equation from the parameters of the fluids. The results
are compared with recent experiments on two-layer systems in which squares at
onset have been reported. 
\end{abstract}

\vspace{0.5cm}


\section{Introduction}
The hexagonal convection cells discovered by B\'enard in his famous
experiments on thin oil layers heated from below \cite{Benard} have become the
trademark of pattern formation in hydrodynamic systems driven slightly out of
equilibrium (see e.g. \cite{CH}). The hundred years of research devoted
to this system have revealed several important insights but also witnessed
several misconceptions. Rayleigh's original theoretical description
\cite{Rayleigh} focusing on buoyancy-driven convection, though indicating a
possible instability mechanism, failed to produce a threshold compatible with
experiment. Not until forty years later was it realized that the 
temperature dependence of the surface tension is the crucial driving force in 
thin layers \cite{Block}. The corresponding linear stability analysis
\cite{Pearson} gave stability thresholds consistent with the experimental
findings, moreover, a subsequent weakly non-linear analysis \cite{ScSe,CL}
produced theoretical support for a sub-critical transition to a hexagonal flow
pattern \cite{Schatz}.

Quite naturally the first theoretical investigations were performed using
simplified models of the experimental situation. The initial assumption of a
flat surface of the liquid was soon relaxed by Scriven and Sternling
\cite{ScSt} and Smith \cite{Smith} who were able to show that surface
deflections give rise to an additional instability appearing at very long
wavelengths. It was only very recently that this 
instability was unambiguously demonstrated in an experiment \cite{Hook} where
it manifests itself as a distortion of the layer thickness with a
characteristic length which is of the order of the lateral extension of the
fluid layer. Being observable only in very shallow liquid layers, the
instability usually results in the formation of dry spots.

Another common simplification is the restriction of the instability mechanism 
to either buoyancy or thermocapillarity \cite{ZeRe,GNP,RBR}, although there
seem to be 
rather few experiments \cite{KoBi,Schatz,Hook} which have been performed in
parameter regions with the ratio between the Rayleigh and the Marangoni number
being sufficiently different from unity. Also, most investigations focussed on
a single layer model in which a lower liquid layer is in contact with a
gaseous upper layer and only the hydrodynamics of the liquid are treated. 
The convection in the gas layer is neglected and the heat exchange between the
layers is often modeled in a phenomenological way using a Biot number, see
e.g. \cite{BrVe}. Even if a genuine two layer model is considered the viscous
stresses and the pressure variations in the gaseous layer are neglected in
order to keep the analysis simple \cite{GNP}. 

On the other hand it has been known for
some time \cite{ZeRe,SiNe} that a system of {\it two superimposed liquids}
displays a much richer behaviour than the single layer models. In particular
the Marangoni instability can be induced by heating from {\it above} such that
buoyancy and thermocapillarity compete rather than enhance each other, a
situation which in single layer systems can only be realized using the rare
case of liquids with anomalous thermocapillary effect in which the surface
tension {\it in}creases with increasing temperature \cite{Bra}. Many additional
features such as oscillatory instabilities \cite{RBR} or transitions from up-
to down-hexagons may be found in systems with two liquid layers. The rich
variety of phenomena occurring in the theoretical analysis of the two-layer
liquid systems results in part from their huge parameter space. A single layer
system is characterized by just three dimensionless parameters; namely, the
Rayleigh number, the Marangoni number and the Prandtl number. The latter is
irrelevant in the linear analysis and the first two are both proportional 
to the temperature difference across the layer. Two layer systems on the other 
hand may easily need ten or more dimensionless parameters for a complete
specification. These numbers include the ratios of the hydrodynamic parameters
of the participating liquids.  

For a long time Marangoni convection in two-liquid-layer systems was an
interesting theoretical problem but too difficult to handle
experimentally. Already Zeren and Reynolds \cite{ZeRe} tried to 
experimentally realize the instability by heating from above which came out of
their theoretical analysis but failed. Very recently, however, experiments
where performed in which the Marangoni instability in 1-2 mm thick
superimposed layers of immiscible liquids was observed \cite{TMM,Anne}. In
particular an instability by heating from above and square patterns at the
onset were reported.  

In the present paper we will investigate theoretically B\'enard-Marangoni
convection in a system of two liquid layers. Building on the linear stability
theory developed in \cite{Hook2} we perform a weakly non-linear analysis in
order to solve the planform selection problem slightly above the linear
stability threshold. To this end the competition between rolls, squares and
hexagons will be discussed. Only perfect patterns will be considered leaving
the question of weakly modulated patterns for future investigations. We will
consistently include buoyancy effects and treat the full hydrodynamics of both
liquids, generalizing in this way various previous treatments
\cite{ScSe,GNP,BrVe,BrLe,Best1,PRLL}. However, we will assume a flat interface
between the two liquids. As will become clear below, interface distortions are
crucial for the long wavelength instability resulting in dry spots but can be
safely neglected when dealing with the finite wavelength instability resulting
in cellular patterns. 

The paper is organized as follows. In section 2 the basic equations are
collected and transformed into a form suitable for the weakly non-linear
analysis. Then the perturbation scheme is set up and the necessary
computational steps are listed. Section 3 deals with the first order of the
perturbation theory which is nothing but the linear stability
analysis. In section 4 the main steps of the nonlinear analysis are
outlined. The solution of the second order problem is relegated to appendix C
and the solvability condition in third order is then formulated to derive the
desired amplitude equation characterizing the planform selection
problem. Section 5 discusses the results obtained for some experimentally
relevant combinations of liquids. Finally section 6 contains a discussion of
the results together with a comparison with experimental findings. 

\section{Basic equations}
\begin{figure}[htb]
  \includegraphics[width=8cm]{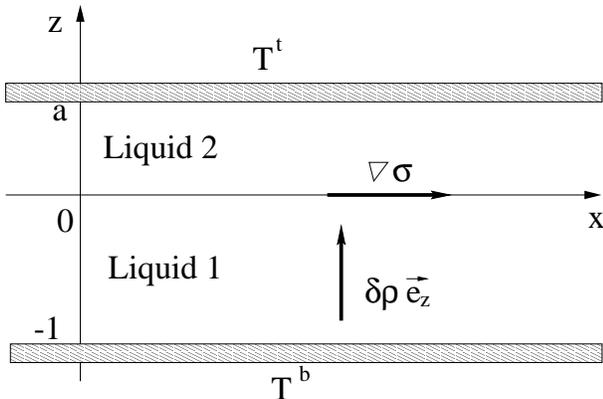}
        \caption{\label{fig1} Sketch of the system under consideration. One
          liquid layer is superposed on another between two horizontally
          infinite, 
          perfectly heat conducting plates. The interface between the liquids
          is assumed to be flat. Convection arises due to buoyancy and the
          temperature dependence of the surface tension.}
\end{figure}

We investigate a system of two layers of immiscible and incompressible liquids
of thickness $h^{(i)}$ with densities $\rho^{(i)}$, kinematic viscosities
$\nu^{(i)}$, coefficients of volume expansion $\alpha^{(i)}$, heat
diffusivities $\chi^{(i)}$, and thermal conductivities $\kappa^{(i)}$ where
the superscript $i=1$ (2) denotes the lower (upper) fluid (see
fig.\ref{fig1}).  
The system is bounded in the vertical direction by two solid, perfectly heat
conducting walls with fixed temperatures $T^{b}$ and $T^{t}$ and is 
infinite in the horizontal directions. The interface between the two fluids is
assumed to be flat and to lie in the $x$-$y$-plane of the coordinate system.  

The hydrodynamics of the two liquids will be described within the
Boussinesq approximation, i.e. we assume that all parameters are
independent of the temperature, except for the densities $\rho^{(i)}$ and the
interface tension $\sigma$. More precisely we use 
$\rho^{(i)}(T)=\rho^{(i)}(T^{b})(1-\alpha^{(i)}(T-T^{b}))$ and 
$\nabla_{\perp}\sigma=d\sigma/dT\, \nabla_{\perp} T$ with constant
$\alpha^{(i)}$ and $d\sigma/dT$. Neglecting heat production due to 
viscosity, the basic equations describing the system are the continuity
equations 
\begin{equation}\label{cont}
\na \ve^{(i)}=0\quad, 
\end{equation}
the Navier-Stokes equations
\begin{equation}\label{iniNSE}
\pa_t\ve^{(i)} + (\ve^{(i)} \na)\ve^{(i)}=-\frac{1}{\rho^{(i)}}\na p^{(i)} -
 g(1-\alpha^{(i)}(T^{(i)}-T^b))\ez +\nu^{(i)}\Delta\ve^{(i)}\quad,
\end{equation}
and the equations of heat conduction 
\begin{equation}\label{heateq}
\pa_t T^{(i)} + (\ve^{(i)} \na)T^{(i)}=\chi^{(i)}\Delta T^{(i)}\quad.
\end{equation}
Here $\ez$ denotes the unit vector in the vertical direction and
$g$ is the acceleration due to gravity. 

These equations are completed by the boundary conditions
\begin{equation}\label{bcb}
  \ve^{(1)}=0\quad,\quad  T^{(1)}=T^b\quad\text{at}\quad z=-h^{(1)}
     \quad,
\end{equation}
and
\begin{equation}\label{bct}
  \ve^{(2)}=0\quad,\quad  T^{(2)}=T^t\quad\text{at}\quad z=h^{(2)}
     \quad,
\end{equation}
at bottom and top respectively and 
\begin{align}\label{bci}
\ve^{(1)}=\ve^{(2)}\quad,\quad T^{(1)}=T^{(2)}\quad,\quad
   \kappa^{(1)}\pa_z T^{(1)}=\kappa^{(2)}\pa_z T^{(2)}\nonumber\\
\left[(\sigma^{(2)}-\sigma^{(1)})\ez\right]_{\perp}
    =-\frac{d\sigma}{d T}\na_{\perp} T \quad,
\quad\ve^{(1)}_z&=\ve^{(2)}_z=0\quad,\qquad\text{at}\quad z=0 \quad,
\end{align}
expressing the continuity of the velocities, temperatures and heat fluxes
respectively as well as the balance of tangential stresses at the interface.
The $\sigma^{(i)}$ denote the stress tensors in the liquids and the subscript
$\perp$ describes the projection to the $x$-$y$-plane.
In accordance with our assumption of a flat interface
between the liquids the condition for the continuity of the normal stress at
the interface is replaced by the requirement that the perpendicular components
of the flow velocities must vanish. This is expressed by the last equation in
(\ref{bci}) .

Introducing $h^{(1)}$, $(h^{(1)})^2/\chi^{(1)}$, $\chi^{(1)}/h^{(1)}$ and 
$\rho^{(1)}\nu^{(1)}\chi^{(1)}/(h^{(1)})^2$ as units for length, time,
velocity, and pressure respectively we find for the velocities 
$\ve=(u,v,w) \,(\Ve=(U,V,W))$ and the appropriately normalized {\it deviations} 
$\te\,(\Te)$ of the temperatures from their static profiles in the lower
(upper) liquid the equations:
\begin{align}
\frac{1}{Pr}(\pa_t\ve + (\ve \na)\ve)&
    =-\na \tilde{p} + R\,\te\,\ez + \Delta \ve\\
  \pa_t\te +(\ve \na)\te&= w + \Delta\,\te\\
\frac{1}{Pr}(\pa_t\Ve + (\Ve \na)\Ve)&
    =-\na \tilde{P} + \alpha\,R\,\Te\,\ez + \nu\Delta\Ve\\
  \pa_t\Te +(\Ve \na)\Te&= \frac{1}{\ka}\, W + \chi \Delta\Te\quad,
\end{align}
where the pressure fields $\tilde{p}$ and $\tilde{P}$ in the lower and the
upper liquid differ from $p^{(1)}$ and $p^{(2)}$ respectively only by 
trivial contributions stemming from the buoyancy terms.
The boundary conditions acquire the form
\begin{equation}\label{bcb1}
  \ve=0\quad,\quad  \te=0\quad\text{at}\quad z=-1
     \quad,
\end{equation}
\begin{equation}\label{bct1}
  \Ve=0\quad,\quad  \Te=0\quad\text{at}\quad z=a
     \quad,
\end{equation}
and 
\begin{align}\label{bci1}
\ve_{\perp}=\Ve_{\perp}\,,\, w&=W=0\,,\, \te=\Te\,,\,
\pa_z \te=\ka\pa_z \Te\,,\,\pa^2_z w-\eta \pa^2_z W=M \Delta_{\perp} \te
             \,,\quad\text{at}\, z=0\,,
\end{align}
where in the last equation the continuity equation was used. Moreover 
the following parameters have been introduced:
\begin{equation}\label{scaling}
a=\frac{h^{(2)}}{h^{(1)}}\;,\;\alpha=\frac{\alpha^{(2)}}{\alpha^{(1)}}\;,\;
\nu=\frac{\nu^{(2)}}{\nu^{(1)}}\;,\;\eta=\nu\frac{\rho^{(2)}}{\rho^{(1)}}\;,\;
\ka=\frac{\ka^{(2)}}{\ka^{(1)}}\;,\;\chi=\frac{\chi^{(2)}}{\chi^{(1)}}\;,
\end{equation}
as well as the Prandtl-number $Pr=\nu^{(1)}/\chi^{(1)}$, the Rayleigh-number
\begin{equation}
  R=\frac{\alpha^{(1)} g (h^{(1)})^3}{\nu^{(1)} \chi^{(1)}}
    \frac{\ka}{a+\ka}(T^b-T^t)\quad,
\end{equation}
and the Marangoni-number
\begin{equation}\label{defM}
  M=-\frac{d\sigma}{d T}\frac{h^{(1)}}{\nu^{(1)}\rho^{(1)}\chi^{(1)}}
      \frac{\ka}{a+\ka}(T^b-T^t)\quad.
\end{equation}
For the Rayleigh- and Marangoni-number we have chosen the standard expressions 
corresponding to the lower liquid. The respective numbers for the upper liquid
are then given by
\begin{equation}
  R^{(2)}=\frac{\alpha a^4}{\nu\chi\kappa} R\quad\text{and}\quad
  M^{(2)}=\frac{a^2}{\chi\eta\kappa} M
\end{equation}
respectively. 

The ratio  between the Rayleigh and Marangoni numbers determines whether the
occurring instability is predominantly driven by buoyancy or by surface
tension. Experimentally both parameters are varied simultaneously since they
are both proportional to the temperature difference $T^b-T^t$. We will
therefore replace $R$ by $c M$ with the temperature independent constant
\begin{equation}
  c=\frac{R}{M}=-\frac{\alpha^{(1)} g (h^{(1)})^2}{d\sigma/d T}
\end{equation}
specifying the experimental setup. In this way both buoyancy and surface
tension are included in a consistent way. We assume that $d\sigma/d T<0$ as is 
the case for most systems of two liquids such that $c>0$. Note that both
the situation of heating from below and heating from above are described 
with the latter case corresponding to $M<0$. 

The set of equations may be simplified by standard manipulations.
Taking twice the curl of the Navier-Stokes equations, using the continuity
equations, and projecting onto $\ez$ we get the following basic set of
equations for the $z$-components of the velocities and the temperature fields:
\begin{align}
\Delta^2 w + c M \Delta_{\perp}\te&=\frac{1}{Pr}\left[\pa_t\Delta w
     -\pa_z(\na_{\perp}(\ve\na)\ve_{\perp})+\Delta_{\perp}(\ve\na)w\right]
              \label{basiceq1}\\
w+\Delta \te&=\pa_t \te+(\ve\na)\te\label{basiceq2}\\
\nu\Delta^2 W + \alpha c M \Delta_{\perp}\Te&=\frac{1}{Pr}\left[\pa_t\Delta W-
     \pa_z(\na_{\perp}(\Ve\na)\Ve_{\perp})+\Delta_{\perp}(\Ve\na)W\right]
              \label{basiceq3}\\
\frac{1}{\ka}W+\chi\Delta\Te&=\pa_t \Te+(\Ve\na)\Te\label{basiceq4}
\end{align}
together with the boundary conditions
\begin{align}
w&=\pa_z w=\te=0\quad\text{at}\quad z=-1\label{basicbc1}\\
w&=W=0\,,\,\pa_z w=\pa_z W\,,\,\te=\Te\,,\,\pa_z\te=\ka\pa_z\Te\,,\,
\pa^2_z w-\eta\pa^2_z W=M\Delta_{\perp}\te \quad\text{at}\quad z=0
   \label{basicbc2}\\
W&=\pa_z W=\Te=0\quad\text{at}\quad z=a\label{basicbc3}
\end{align}
In order to investigate the planform selection problem we will derive third
order amplitude equations for the slow time variation of the amplitudes of
different unstable modes. Similar to the case of the Rayleigh-B\'enard
instability \cite{CH} the no-slip boundary conditions at top and bottom
suppress the 
vertical vorticity, i.e. $(\na\times\ve)\ez=(\na\times\Ve)\ez=0$, and
therefore we do not expect problems due to a coupling to a slowly varying mean
flow \cite{ZiSi} up to this order. From the solution of
(\ref{basiceq1})-(\ref{basiceq4}) we hence obtain $w,\te,W$ and $\Te$. Using
the continuity equations and the absence of vertical vorticity allows to
determine $u,v$ and $U,V$ and finally the pressure fields follow from the
Navier-Stokes equations. 

It is convenient to write the above equations in the form
\begin{equation}\label{h1}
  L \ph = \cal{T}(\ph)+\cal{N}(\ph,\ph)
\end{equation}
with the state vector 
\begin{equation}\label{defphi}
  \ph=\begin{pmatrix} w\\ \te \\ W \\ \Te \end{pmatrix}\quad,
\end{equation}
and the linear operator $L$ defined by 
\begin{equation}\label{defL}
  L=\begin{pmatrix}
     \Delta^2  & c M \Delta_{\perp}& 0 & 0\\
         1     &      \Delta       & 0 & 0\\
         0     &       0           & \nu\Delta^2 &\alpha c M \Delta_{\perp}\\ 
         0     &       0           &\frac{1}{\ka} & \chi\Delta\\
    \end{pmatrix}\quad,
\end{equation}
and the boundary conditions (\ref{basicbc1})-(\ref{basicbc3}). $\cal{T}(\ph)$
denotes the time dependent terms and $\cal{N}(\ph,\ph)$ describes the
quadratic nonlinearity in (\ref{basiceq1})-(\ref{basiceq4}). We will solve
(\ref{h1}) perturbatively using the ans\"atze  
\begin{align}
  \ph&=\eps\ph_0 + \eps^2 \ph_1 + \eps^3 \ph_2+\dots\label{pertans1}\\
    M&=M_c + \eps M_1 + \eps^2 M_2+\dots\label{pertans2}\\
\pa_t&=i\omega\qquad\quad\;\; +\eps^2 \pa_\tau+\dots \label{pertans3}
\end{align}
with a small parameter $\eps$. In the case of a static instability we have
$\omega=0$ whereas for an oscillatory instability $\omega\neq 0$ gives the
frequency of oscillation of the unstable mode. Using the perturbation
expansion specified above we consider a situation slightly above the threshold
$M_c$ of the linear instability, where the amplitude of the unstable modes can
still be considered to be small. 
Plugging (\ref{pertans1})-(\ref{pertans3}) into (\ref{h1}), taking into
account that (\ref{pertans2}) implies an expansion 
\begin{equation}\label{Lexp}
L=L_0+\eps L_1+\eps^2 L_2+\dots
\end{equation}
for the linear operator and matching powers of $\eps$ the non-linear problem
transforms into a sequence of linear equations of the form 
\begin{align}
  L_0 \ph_0 &=0\label{hier1}\\
  L_0 \ph_1 &=-L_1 \ph_0 + {\cal N}(\ph_0,\ph_0)\label{hier2}\\
  L_0 \ph_2 &=-L_2 \ph_0 -L_1\ph_1 +{\cal T}(\ph_0) + {\cal N}(\ph_1,\ph_0)
               + {\cal N}(\ph_0,\ph_1)\label{hier3}\quad.
\end{align}
The first line is just the linear stability problem. The condition for
non-trivial solutions $\ph_0$ of this equation makes $L_0$ singular and yields
the critical value  $M_c$ of the bifurcation parameter $M$. From the translation invariance in the $x$-$y$-plane we know that $\ph_0$ is
of the form 
\begin{equation}
\ph_0=\ph_0(z)\exp(i\bk \br-i\omega t)
\end{equation}
where $\br=(x,y)$ and $\bk=(k_x,k_y)$ 
are two-dimensional vectors. There is a critical value $M_c(k)$ of the
bifurcation parameter for all values of $|\bk|=k$ and minimizing $M_c(k)$ in
$k$ gives the wavenumber $k_c$ of the first unstable mode together with the
critical Marangoni number $M_c=M_c(k_c)$.

The remaining equations in the hierarchy starting with (\ref{hier2}) all
involve the {\it very same} 
singular operator $L_0$ but are {\it in}homogeneous. Consequently the 
perturbation expansion makes sense only if the inhomogeneities are
perpendicular to the zero eigenfunction of the adjoint operator $L^+_0$ of
$L_0$.  

\begin{figure}[htb]
  \includegraphics[width=7cm]{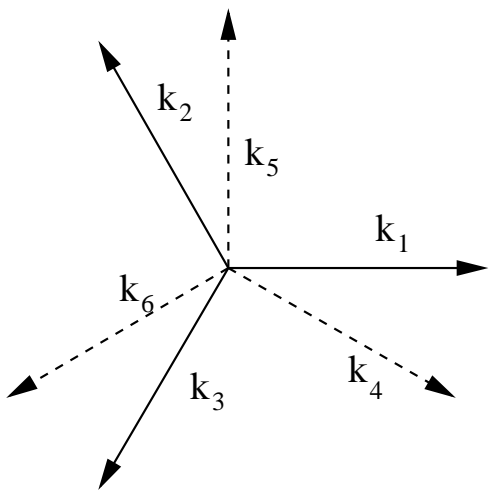}
        \caption{\label{fig2} Relative orientation of the two-dimensional
          wave vectors appearing in the ansatz (\ref{formph0}). The two triads 
          $\bk_1,\bk_2,\bk_3$ and  $\bk_4,\bk_5,\bk_6$ of wave vectors with
          $\bk_5$ perpendicular to $\bk_1$ allow to describe rolls as well as
          squares and hexagons by different values for the amplitudes $A_n$ in 
          (\ref{formph0}).}
\end{figure}

In order to address the planform selection problem within the perturbation
approach sketched above the form of $\ph_0$ must be sufficiently general and
in particular must include the different planforms observed in the
experiment. We will discuss the planform selection problem only for the case
of the static instability leaving the investigation of the oscillatory
instability to future work. It is then sufficient to use for $\ph_0$ the form 
\begin{equation}\label{formph0}
  \ph_0=\ph_0(z)\left[\sum_{n=1}^6 A_n(\tau) e^{i\bk_n \br}+\text{c.c.}\right]
\end{equation}
with the six two-dimensional vectors $\bk_n$ obeying
$|\bk_n|=k_c$ and $\bk_1+\bk_2+\bk_3=0$, $\bk_4+\bk_5+\bk_6=0$, as well as 
$\bk_1\bk_5=0$ (see fig.\ref{fig2}). Depending on the values of the amplitudes
$A_n$ this form describes rolls (e.g. $A_1=A, A_n=0$ for all $n>1$), 
squares (e.g. $A_1=A_5=A, A_n=0$ else) and hexagons 
(e.g. $A_1=A_2=A_3=A$, $A_n=0$ for $n>3$). 

Using this form we find from the solvability conditions of (\ref{hier2}) and 
(\ref{hier3}) an equation describing the time evolution
of the scaled amplitudes $\tilde{A_n}=\eps A_n$. As is well known \cite{CH}
the {\it general} form of 
this amplitude equation already follows from the symmetries of the
problem. For the present situation it is given by
\begin{equation}\label{ampl}
\pa_t \tilde{A}_1=\epsilon \tilde{A}_1 +\gamma \tilde{A}_2^* \tilde{A}_3^* 
  -\left[|\tilde{A}_1|^2 +g_h(|\tilde{A}_2|^2+|\tilde{A}_3|^2) + 
         g_t(|\tilde{A}_4|^2+|\tilde{A}_6|^2) 
        +g_n |\tilde{A}_5|^2\right] \tilde{A}_1
\end{equation}
with the super-criticality parameter
\begin{equation}\label{defeps}
  \epsilon=\frac{M-M_c}{M_c}\quad.
\end{equation}
Similar equations for the other amplitudes follow from 
permutation and complex conjugation. The terms included in these equations are
the only ones up to third order which are invariant under the transformation 
$A_n\mapsto A_n\exp(i\bk_n \br_0)$ corresponding to a translation by $\br_0$
in the $x$-$y$-plane. Moreover due to the isotropy in the $x$-$y$-plane the
coupling coefficients between the different terms in (\ref{formph0}) may only
depend on the angle between the corresponding wave vectors. 

A well known linear stability analysis of the various fix points of
(\ref{ampl})  
yields the stability regions of the different planforms as functions of the
parameters $\epsilon, \gamma, g_h, g_t, g_n$ \cite{Cil}. The remaining problem
is hence to use the perturbation expansion described above to express these
coefficients of the amplitude equation in terms of the hydrodynamic parameters 
of the problem. To this end the following well-known program has to be carried
through: 
\begin{itemize}
\item Calculate $M_c(k)$ from the linear problem and determine 
  $k_c=\text{argmin}\, M_c(k)$ and $M_c=M_c(k_c)$. 
\item Determine the adjoint operator $L^+_0$ of $L_0$ and its zero 
  eigenfunction $\bar{\ph}_0$.
\item Calculate the inhomogeneity of the $O(\eps^2)$-equation (\ref{hier2})
  and apply the solvability condition to this order.
\item Solve the $O(\eps^2)$-equation (\ref{hier2}) to determine $\ph_1$.
\item Calculate the inhomogeneity of the $O(\eps^3)$-equation (\ref{hier3}) 
  (only terms proportional to $\exp{i(\bk_1 \br)}$ are necessary)
\item Combine the solvability conditions at order $O(\eps^2)$ and $O(\eps^3)$
  to derive (\ref{ampl}) and extract the expressions for the parameters 
  $\gamma, g_h, g_t, g_n$.
\end{itemize}

\section{The linear problem}

We first solve the $O(\eps)$ problem (\ref{hier1}), which is equivalent to the 
linear stability analysis. Putting
\mbox{$\ph_0=\ph_0(z)\exp(i\bk\br-i\omega t)$} and using the ans\"atze
\begin{alignat}{2}
  w_0(z),\te_0(z) &\sim\exp(\la z)\qquad&\qquad 
  W_0(z),\Te_0(z) &\sim\exp(\La z)
\end{alignat}
we find 
\begin{alignat}{2}\label{eqla}
  (\la^2-k_c^2)(\la^2-k_c^2+\frac{i\omega}{Pr})(\la^2-k_c^2+i\omega)
           &=-cMk_c^2\quad&\quad 
  (\La^2-k_c^2)(\La^2-k_c^2+\frac{i\omega}{\nu Pr})
(\La^2-k_c^2\frac{i\omega}{\chi})&=-\frac{\alpha}{\nu\kappa\chi}cMk_c^2\quad.
\end{alignat}
We therefore obtain six different values for $\la_i$ and $\La_i$. It is
convenient to define $\la_i=\La_{i-6}$ for $i=7,\dots,12$ and to write 
\begin{alignat}{2}
  w_0(z)&=\sum_{i=1}^6 w_{0i}\,e^{\la_i z}\qquad&\qquad 
\te_0(z)&=-\sum_{i=1}^6 
         \frac{w_{0i}}{\la_i^2-k_c^2+i\omega}\,e^{\la_i z}\label{h3}\\
  W_0(z)&=\sum_{i=7}^{12} w_{0i}\,e^{\la_i z}\qquad&\qquad 
\Te_0(z)&=-\frac{1}{\kappa\chi} \sum_{i=7}^{12} 
    \frac{w_{0i}}{\la_i^2-k_c^2+\frac{i\omega}{\chi}}\,e^{\la_i z}\label{h2}
\end{alignat}
The boundary conditions (\ref{basicbc1})-(\ref{basicbc3}) give then rise to a
homogeneous system of linear equations for the 12 unknowns $w_{0i}$. In order
to get a non-trivial solution the determinant of the coefficient matrix 
${\cal A}$ must vanish. The conditions for the real and the imaginary part of
det${\cal A}$ yield the desired functions $M_c(k;\para)$ and
$\omega_c(k,\para)$ where $\para=(a,\alpha,\kappa,\chi,\nu,\eta,c,Pr)$ stands
for the vector of parameters in the problem. 
\begin{figure}[htb]
  \includegraphics[width=10cm]{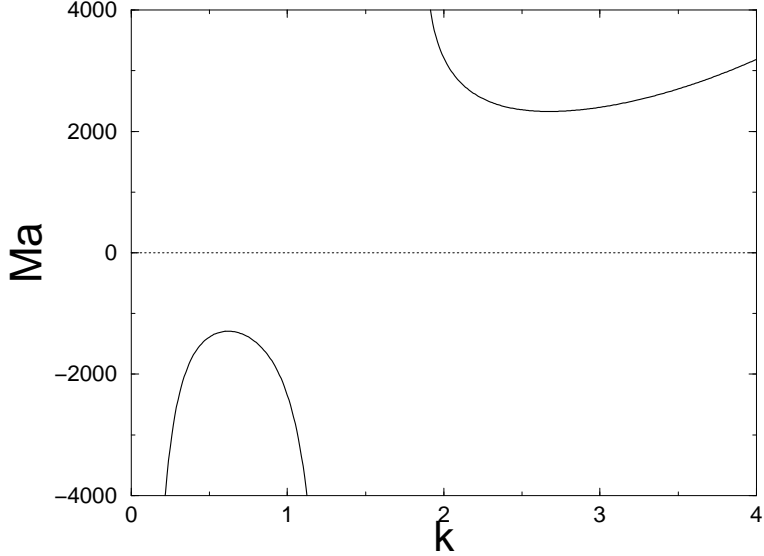}
        \caption{\label{figlin} Dispersion relation $M_c(k)$ as resulting from 
          the linear stability analysis for the hydrodynamic parameters of
          setup 2 listed in appendix A. The system shows an instability when
          heated from below as well as one when heated from above.}
\end{figure}

A typical result for a static instability is shown in fig.\ref{figlin}
displaying  the dispersion curve resulting from the numerical analysis of
det${\cal A}=0$ for $\omega=0$ using the parameters 
of setup 2 listed in appendix A. As can be seen from the 
figure in this system one may have an instability by heating from below
($M>0$) as well as when heating from above ($M<0$). In fig.\ref{figlin2} the
results of the present approach for the setups 1 and 5 of appendix A 
are compared
with those resulting from the full linear stability analysis including surface 
deflections as considered in \cite{Hook2}. As is clearly seen in the region of
the pattern forming instability $k\cong 1...3.5$ the two curves are almost
identical with differences showing up only for small wave numbers $k\ll 1$. 
Within the linear theory the surface deflections for unstable modes 
corresponding to the planform selection problem may therefore safely be
neglected and we expect that this is also a good approximation for 
the weakly non-linear regime. 

\begin{figure}[htb]
  \includegraphics[width=10cm]{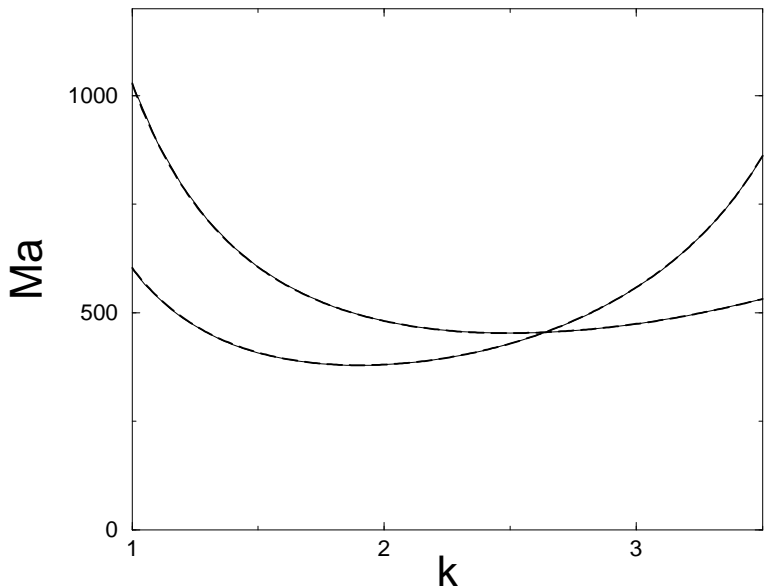}
        \caption{\label{figlin2} Comparison of the dispersion relation
          $M_c(k)$ as resulting from the present analysis assuming a flat 
          interface between the liquids (thin full lines) with the results of
          the complete linear analysis of \protect\cite{Hook2} including
          surface deflections (dashed lines). The lower right curves are for
          setup 1, the others for setup 5 specified in appendix A.}
\end{figure}
Having obtained the dispersion relation we calculate $k_c$ by minimizing
$M_c(k)$ and determine the critical Marangoni and Rayleigh numbers of both
fluids as well as the temperature difference across both layers at the
instability. The results for the setups under consideration are summarized in
the upper part of table 1. 

\begin{figure}[htb]
  \includegraphics[width=10cm]{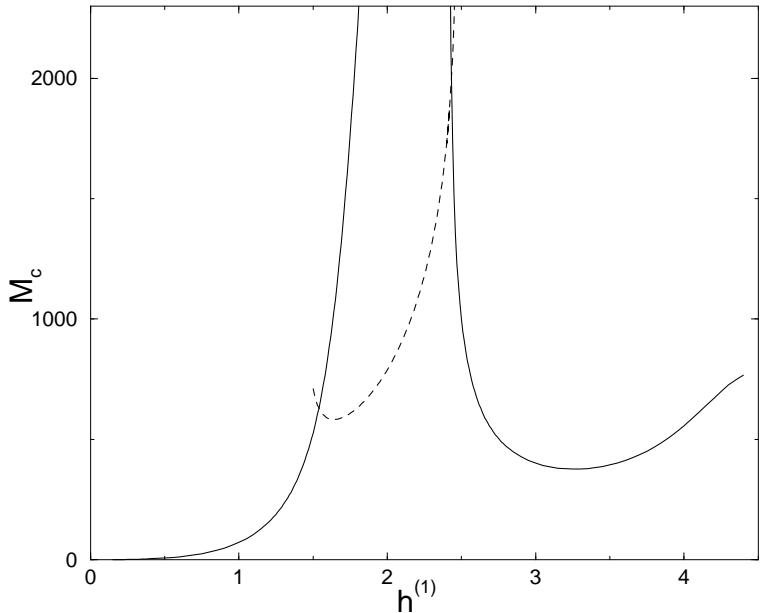}
        \caption{\label{ahlin} Critical Marangoni number for heating from
          below a system with parameters as specified in setup 3 of appendix A
          and total depth 4.5 mm as a function of the bottom layer thickness
          $h^{(1)}$. Note that both $M$ and $k$ are scaled with $h^{(1)}$
          (cf. (\ref{scaling}) and (\ref{defM})).}
\end{figure}

From all the parameters of the system the depth ratio $a$ is the only one
which may be easily varied in the experiments. For the parameters of setup 3
and a total depth of 4.5 mm we have calculated the critical Marangoni number
and the critical wave number modulus as a function of the thickness $h^{(1)}$
of the bottom layer restricting ourselves to the case of heating from below
but including the possibility of an oscillatory instability. 
The results are displayed in fig.\ref{ahlin}. For values of $h^{(1)}$ between 
1.5 and 2.5 an oscillatory instability precedes the static one which would
occur at unusually large Marangoni numbers only. A similar oscillatory
instability was also found for a two-layer system in which the Marangoni
effect was neglected and pure buoyancy-driven convection was considered, and
an intuitive interpretation as a periodic change between viscous and thermal
coupling of the flow fields at the interface was given \cite{RBR}.
The oscillatory instability was also detected in the experiment using setup 3
with $h^{(1)}\cong 1.8$mm and the experimental values for the critical
Marangoni number and the wavelength of the oscillatory mode are in good
agreement with the theory \cite{Anne}.  

Knowing the critical value of $M$ we can now also determine the coefficients
of the eigenvector corresponding to the zero eigenvalue. This fixes the
functions $w_0(z),\te_0(z),W_0(z)$ and $\Te_0(z)$ up to an overall constant
and completes the determination of $\ph_0$. 

Finally we have to consider the adjoint problem and to calculate its zero
eigenfunction $\bar{\ph}_0$ where we again restrict ourselves to the
stationary instability. The adjoint operator $L^+$ is determined in
appendix B. The calculation of its eigenfunction to the eigenvalue zero is
very similar to the determination of $\ph_0$ described above. We find that it
is of the form $\bar{\ph_0}\exp(i\bk_n\br)$ where the components of
$\bar{\ph_0}$ may be written as 
\begin{alignat}{2}
  \bar{w}_0(z)&=\sum_{i=1}^6 \bar{w}_{0i}\,e^{\la_i z}\qquad&\qquad 
\bar{\te}_0(z)&=cMk_c^2\sum_{i=1}^6 
   \frac{\bar{w}_{0i}}{\la_i^2-k_c^2}\,e^{\la_i z}\\
 \bar{W}_0(z)&=\sum_{i=7}^{12} \bar{w}_{0i}\,e^{\la_i z}\qquad&\qquad 
\bar{\Te}_0(z)&=\frac{\alpha c M k_c^2}{\chi}
   \sum_{i=7}^{12} \frac{\bar{w}_{0i}}{\la_i^2-k_c^2}\,e^{\la_i z}
\end{alignat}
with the same parameters $\la_i$ as determined by (\ref{eqla}) with
$\omega=0$. The boundary conditions give again rise to a $12\times12$ system
of linear homogeneous  
equations for the coefficients $\bar{w}_{0i}$. As before the condition for a
non-trivial solution is a vanishing determinant of the corresponding
matrix. Note, however, that there is now no parameter to adjust! The deviation
of the smallest eigenvalue of the matrix found in the numerical calculation
from zero gives therefore a valuable hint on the accuracy of the numerical
procedure employed.

\section{The nonlinear analysis}

The solution of the planform
selection problem requires the treatment of the nonlinear
interaction between different unstable modes. To include nonlinear terms
up to the third order in the amplitudes $A_n$ introduced in (\ref{formph0}) we
have first to solve (\ref{hier2}). The general procedure is standard, some
intermediate steps are sketched in appendix C. 
Using this solution we are in the position to calculate the terms appearing on
the right hand side of (\ref{hier3}). We do not have to  
solve this equation, but only need to know the solvability condition at this
order. Due to the $x$-$y$-integrals in (\ref{defscalprod}) and the
$\br$-dependence of $\bar{\ph_0}$ only terms proportional to
$\exp(\pm i\bk_n\br)$  
give rise to non-trivial contributions to the solvability condition. In fact
it is sufficient to focus on terms proportional to $\exp(i\bk_1\br)$ since
these finally give rise to an amplitude equation of the form 
(\ref{ampl}) for $A_1$. Equivalent equations for the other amplitudes of the
ansatz (\ref{formph0}) follow then from permutation and complex conjugation.  

In order to collect the relevant terms we first realize that there are
contributions  
\begin{equation}
A_1\,e^{i\bk_1\br}
\begin{pmatrix} c\,M_2\,k_c^2\te_0\\ 0\\ \alpha c\,M_2\,k_c^2\Te_0 \\ 0 \\ 
                 -M_2\,k_c^2 \te_0|_{z=0}\end{pmatrix}\quad,\quad
A_2^*A_3^* \,e^{i\bk_1\br}
\begin{pmatrix} c\,M_1\,k_c^2\te_1\\ 0\\ \alpha c\,M_1\,k_c^2\Te_1 \\ 0 \\ 
                 -M_1\,k_c^2 \te_1|_{z=0}\end{pmatrix}\quad,\quad
\pa_{\tau}A_1\,e^{i\bk_1\br}
\begin{pmatrix} \frac{1}{Pr}(w_0''-k_c^2 w_0)\\ \te_0\\ 
                \frac{1}{Pr}(W_0''-k_c^2 W_0)\\ \Te_0 \\ 0\end{pmatrix}\quad,
\end{equation}
originating from the terms $-L_2\ph_0$, $-L_1\ph_1$, and ${\cal T}(\ph_0)$
respectively in (\ref{hier3}). Here $\te_1$ and $\Te_1$ denote the solutions
obtained in the last section for the resonant term. 

The contributions proportional to $\exp(i\bk_1\br)$ from the last two terms in
(\ref{hier3}) arise from combinations between $\ph_0\sim\exp(i\bq\br)$ and
$\ph_1\sim\exp(i\bp\br)$ with $\bq+\bp=\bk_1$. From the continuity equation,
$\na\ve=0$, and the absence of vertical vorticity, $(\na\times\ve)\ez=0$, we
find  
\begin{equation}
\ve_{0\perp}=e^{i\bq\br}\,\frac{i\bq}{\bq^2}\,\pa_z w_0\qquad,\qquad
\ve_{1\perp}=e^{i\bp\br}\,\frac{i\bp}{\bp^2}\,\pa_z w_1\quad,
\end{equation}
which gives rise to
\begin{multline}
-\pa_z(\na_{\perp}(\ve_0\na)\ve_{1\perp})+\Delta_{\perp}(\ve_0\na) w_1
-\pa_z(\na_{\perp}(\ve_1\na)\ve_{0\perp})+\Delta_{\perp}(\ve_1\na) w_0=\\
e^{i\bk_1\br}\!\!\left[\frac{\bk_1\bq}{\bq^2}w_0''' w_1 
  \!+\!(\frac{\bk_1\bq}{\bq^2}-\frac{\bq\bp}{\bq^2\bp^2}k_c^2) w_0'' w_1'
  \!+\!(\frac{\bk_1\bp}{\bp^2}-\frac{\bq\bp}{\bq^2\bp^2}k_c^2) w_0' w_1''
  \!+\!\frac{\bk_1\bp}{\bp^2} w_0 w_1'''
+k_c^2\left( (\frac{\bq\bp}{\bq^2}-1)\,w_0' w_1 
            \!+\!(\frac{\bq\bp}{\bp^2}-1)\,w_0 w_1'\right)\right]\nonumber
\end{multline}
and 
\begin{equation}
(\ve_0\na)\te_1 + (\ve_1\na)\te_0=
e^{i\bk_1\br}\left[-\frac{\bq\bp}{\bq^2}\,w_0'\te_1
                   -\frac{\bq\bp}{\bp^2}\,w_1'\te_0 
         +w_0\te_1'+w_1\te_0'\right]\quad.\nonumber
\end{equation}
With the help of these relations it is now easy to determine the remaining
terms proportional to $\exp(i\bk_1\br)$ from all the possible combinations for 
$\bq$ and $\bp$ and the corresponding results for $\ph_1$ calculated in
appendix C. 

Using the scalar product (\ref{defscalprod}) and the result for $\bar{\ph_0}$, 
the solvability condition at order $O(\eps^3)$ can be formulated. It
contains a term proportional to $M_1 A_2^*A_3^*$ which by eliminating $M_1$
using the solvability condition (\ref{F2}) at order $O(\eps^2)$ is transformed
into terms 
proportional to $|A_2|^2A_1$ and  $|A_3|^2A_1$. We then multiply the
solvability condition at order $O(\eps^2)$ by $\eps^2$ and the one at order
$O(\eps^3)$ by $\eps^3$ and add them together. Observing 
$\eps M_1+\eps^2 M_2=M-M_c$, returning to the original time by using 
$\eps^2\pa_{\tau}=\pa_t$ and introducing the scaled amplitudes
$\tilde{A}_n=\eps A_n$ we eventually end up with an amplitude equation of the
form (\ref{ampl}) with explicit expressions for the parameters $\gamma, g_h,
g_t$ and $g_n$.

\section{Results}

The expressions for $\gamma, g_h, g_t$ and $g_n$ are rather long and
will not be displayed. Moreover, due to the large number of parameters in the
two liquid system it is more appropriate to analyze some experimentally
relevant parameter combinations rather than to display cross sections along
some direction of the parameter space. For the five experimental setups
specified in appendix A the results of the non-linear analysis are summarized
in the lower part of table 1.  

In order to finally address the planform selection problem we note that from
the linear stability analysis of the roll, square and hexagon solutions of
the amplitude equation (\ref{ampl}) it is well known \cite{Cil,BrVe} that:
\begin{itemize}
\item rolls are stable if $g_h>1$, $g_t>1$, $g_n>1$, and
  $\epsilon>\frac{\gamma^2}{(1-g_h)^2}$,
\item squares are stable if $1+g_n<g_h+g_t$, $|g_n|<1$, and 
  $\epsilon>\frac{\gamma^2(1+g_n)}{(1+g_n-g_h-g_t)^2}$,
\item hexagons are stable if $1+2g_h>0$,
  $\epsilon>\epsilon_h=-\frac{\gamma^2}{4(1+2g_h)}$, either $g_h<1$ or
  $\epsilon<\epsilon_{htr}=\frac{\gamma^2(2+g_h)}{(1-g_h)^2}$, 
  and either $1+2g_h<g_n+2g_t$ or $\epsilon<\epsilon_{hts}=
  \frac{\gamma^2(g_n+2g_t)}{(1+2g_h-g_n-2g_t)^2}$.  
\end{itemize}
In addition to the special values of $\epsilon$ defined in the last point
above we have also 
included in table 1 the amplitude $A_h$ of the pattern at onset. The hexagon
pattern appears through a backward bifurcation which strictly speaking
invalidates our perturbation ansatz (\ref{pertans1}). However, the
interval of sub-critical hexagons as well as the amplitude of the pattern at
onset are for all investigated setups rather small such that the ansatz is
still a good approximation for what really happens. 

Except for setup 3 when heated from below we always find $g_n<1$ excluding the 
possibility of stable rolls within the framework of 
our weakly non-linear analysis. For all setups we get $1+2g_h>0$ which
implies that for hexagons the cubic term is able to stabilize the linear
instability. Moreover for all setups $g_h>1$ and $1+2g_h>g_n+2g_t$ implying 
that the values of $\epsilon_{htr}$ and $\epsilon_{hts}$ give the stability
border for hexagons. Being the result of an expansion in the
amplitude of the unstable modes the numerical values of $\epsilon_{htr}$ and
$\epsilon_{hts}$ are only reliable if they are not too large. If these values
are hopelessly outside the validity of our perturbation approach they are not
displayed in table 1. In all other cases we find  
$\epsilon_{htr}>\epsilon_{hts}$ for setups with $g_n<1$ in accordance with the
fact that rolls are then unstable to squares. The value of 
$\epsilon_{hts}$ is always positive which means that exactly {\it at} onset our
analysis always predicts hexagons as the stable planform and excludes
squares. However, in the cases where $\epsilon_{hts}$ is rather
small (e.g. setup 4 when heated from below) hexagons get very quickly
unstable to squares when passing the stability threshold. 

The sign of $\gamma$ is related to the detailed 
convection pattern of the hexagon planform. For $\gamma>0$ the hexagons in the 
lower fluid are up-hexagons (liquid rises in the center) and the one in the
upper layer are down-hexagons. For $\gamma<0$ the situation is reversed. We do
not know of experimental results concerning this feature for the two liquid
Marangoni problem. \\[.5cm]
\hspace*{.5cm}
\setlength{\tabcolsep}{4mm}
\begin{tabular}{||c||c||c|c||c|c||c|c||c||}\hline
& setup 1&\multicolumn{2}{c||}{setup 2}
&\multicolumn{2}{c||}{setup 3}&\multicolumn{2}{c||}{setup 4}
& setup 5\\[2mm]\hline\hline
$\Delta T$& 0.415 & 4.032 & -3.945& 1.523& -0.256
          & 0.859 & -18.957 & 1.718\\\hline
$k_c$ & 2.495 & 2.745 & 0.714 & 4.3416 & 1.0328
      & 2.377 & 0.861 & 1.901\\\hline
$M$ & 453 & 1919 & -1878 & 1978 & -333
    & 869 & -19188 & 379 \\\hline
$R$ & 676 & 654  & -640  & 669 & -113 
    & 733 & -16168 & 45 \\\hline
$M^{(2)}$& 24.1 & 614 & -601 & 12107 &-2036 
         & 149 & -3284 & 777 \\\hline
$R^{(2)}$& 4.88 & 592 & -579 & 8145 & -1370 
         & 49 & -1091 &  143 \\\hline\hline
$\gamma$ & 0.406 & 0.367& -0.559 & -0.7478 &-0.5428 
         & 0.423 & -0.507 & 0.430 \\\hline
$g_h$& 1.225 & 1.196 & 1.411 & 1.57 & 1.36 
     & 1.188 & 1.417 & 1.377\\\hline
$g_t$& 1.442 & 1.480 & 1.501 & 1.021 & 1.529
     & 1.164 & 1.273 & 1.551\\\hline 
$g_n$& 0.030 & 0.419 & 0.075 & 1.594 & -0.027  
     & -0.355 & -0.050 & 0.628\\\hline\hline
$\epsilon_h$& -0.012 & -0.010 & -0.020 & -0.034 & -0.020 
            & -0.013 & -0.017 & -0.012\\\hline
$A_h$& 0.118 & 0.108 & 0.1462 & 0.180 & 0.146 
     & 0.125 & 0.132 & 0.114\\\hline
$\epsilon_{htr}$& - & - & 6.30 & 6.12 & 7.50 
                       & - & 5.04 & 4.38 \\\hline
$\epsilon_{hts}$& 1.670 & - & 1.734 & 7.922 & 1.848
                       & 0.180 & 0.358 & -\\\hline
\end{tabular}

\vspace{.5cm}
\noindent Table 1: Results for the critical temperature difference $\Delta T$
  over both liquids ($\Delta T>0$ for heating from below, $\Delta T<0$ for
  heating from above), the critical wavenumber $k_c$, the Marangoni and
  Rayleigh numbers of both liquids at onset, the parameters of the amplitude
  equation (\ref{ampl}), the sub-critical threshold $\epsilon_h$ for the
  hexagonal pattern, its amplitude $A_h$ at onset, and the values
  $\epsilon_{htr}$ and $\epsilon_{hts}$ at which the hexagon 
  pattern gets unstable towards the formation of rolls and squares
  respectively. If the numerical values of $\epsilon_{htr}$ and
  $\epsilon_{hts}$ obtained are larger than 10 they are meaningless as result
  of a perturbation expansion in $\epsilon$ and are therefore not displayed.\\

\begin{figure}[htb]
  \includegraphics[width=10cm]{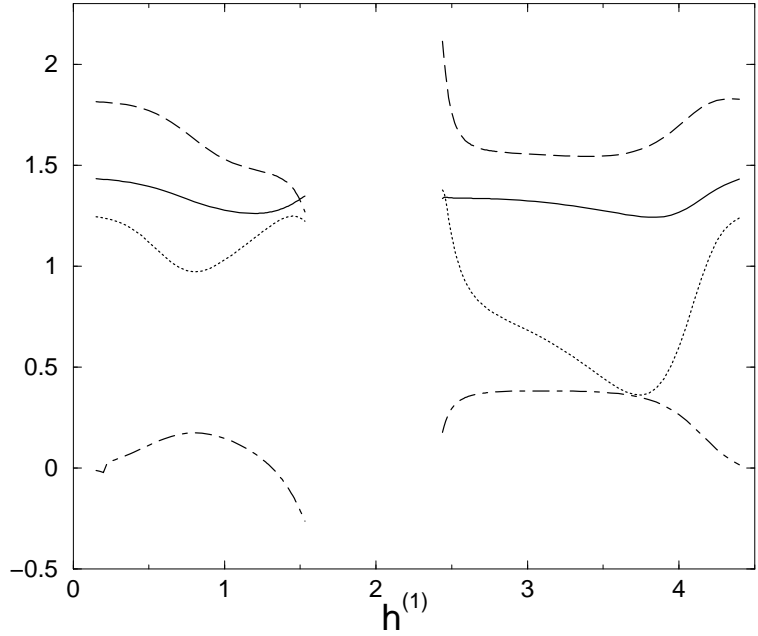}
        \caption{\label{ahnl1} The parameters $g_h$ (full),
          $g_t$ (dashed), $g_n$ (dotted) and $\gamma$ (dashed-dotted) of the
          amplitude equation (\ref{ampl}) as functions of the thickness
          $h^{(1)}$ of the bottom layer for setup 3 with a total layer depth
          of 4.5 mm and heating from below. For 
          1.5 mm$\lesssim h^{(1)}\lesssim 2.5$ mm the oscillatory instability
          precedes the static one.} 
\end{figure}

For the parameters of setup 3 and a total depth of 4.5 mm we have again
scanned the dependence of the results of the non-linear analysis on the
thickness of 
the bottom layer for the case of heating from below. Fig.\ref{ahnl1} shows
the coefficients of the amplitude equation (\ref{ampl}) as functions of
$h^{(1)}$. The most apparent feature is the strong sensitivity of the
coefficients on variations of the depth ratio. In experiments the depth must
therefore be controlled very accurately in order to allow sensible
comparison with the theory. The system under consideration shows a transition
from up to down hexagons when varying the depth ratio as can be seen from the
change of the sign of $\gamma$.

\begin{figure}[htb]
  \includegraphics[width=10cm]{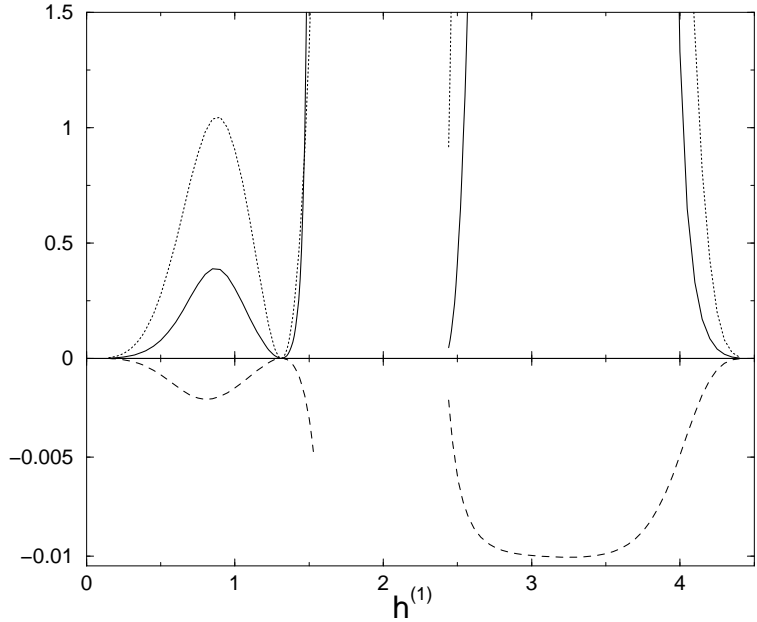}
        \caption{\label{ahnl2} The values of $\eps_{htr}$ for the transition
          from hexagons to rolls (dotted) and $\eps_{hts}$ for the transition
          from hexagons to squares (full) as functions of the thickness
          $h^{(1)}$ of the bottom layer for setup 3 with a total layer depth
          of 4.5 mm and heating from below. Also shown is the value $\eps_h$
          at which hexagons appear sub-critically. Note the different scales
          for positive and negative values at the vertical axis. For 
          1.5 mmm $\lesssim h^{(1)}\lesssim 2.5$ mm the oscillatory instability
          precedes the static one.}
\end{figure}

Finally in fig.\ref{ahnl2} the dependence of $\eps_{htr}$ and $\eps_{hts}$ on
$h^{(1)}$ is displayed. For most values of $h^{(1)}$ we have 
$\eps_{htr}>\eps_{hts}$ and the hexagon pattern becomes unstable to the
formation of squares. However for $h^{(1)}\cong 1.5$ also a secondary
transition to rolls is possible. For some values of the depth ratio we find a
very small $\eps_{hts}$. Since at the same time also the absolute value
of $\eps_h$ is very small implying a small hysteretic window for the formation
of hexagons it is quite conceivable that in these situations in the experiment
the hexagon pattern cannot be observed at all and squares are seen directly at 
onset. 

\section{Discussion}
In the present paper a weakly non-linear analysis for B\'enard-Marangoni
convection in systems of two superimposed liquids has been developed. A
consistent treatment of the full hydrodynamics and heat conduction in both
layers was performed. As crucial simplifying ingredient of our approach we
have used the assumption of an undisturbed interface between the
liquids. Comparison with the complete linear stability analysis including
interface deflections has revealed that this approximation is extremely good
for the pattern forming instability occurring at not too long wavelengths. We
have considered the planform selection problem by determining the relative
stabilities of roll, square and hexagon patterns. To this end the coefficients 
of the appropriate amplitude equation were calculated as functions of the
hydrodynamic parameters by a perturbation theory in the amplitude of the
unstable mode. As is well known \cite{Busse}, this expansion is not rigorous
for the case of a sub-critical bifurcation leading to a finite amplitude
immediately at onset. However, for the parameter combinations 
used we found that the hysteresis, as measured by $\epsilon_h$, is weak and
the results obtained should therefore be rather accurate. 

Explicit numerical results were obtained for five different sets of
experimentally relevant parameters of the fluids. Since the system is
on the one hand characterized by nine dimensionless parameters whereas it is on
the other hand very hard to find two really immiscible fluids to perform the
experiments this seems to be the most sensible way to theoretically
investigate the peculiarities of the system which may also be seen in
experiments. 
For all parameter combinations investigated we predict hexagons at onset in 
agreement with recent experimental findings \cite{Anne}. This shows that
extrapolations from previous results on liquid-gas systems \cite{GNP} to the
two liquid layer system which gave arguments in favour of squares directly at
onset are potentially dangerous and the full hydrodynamics of both layers has
to be taken into account. Moreover in most cases rolls were found to be
unstable to squares for all values of the super-criticality parameter
$\epsilon$. In particular for the parameter values of the experiments
described in \cite{TMM} we do not find stable rolls in contrast to the
secondary transition from squares to rolls reported for this case.

The hexagonal pattern gets unstable to squares at a positive value
$\epsilon_{hts}$ of the super-criticality parameter. For different experimental 
setups the values of $\epsilon_{hts}$ differ substantially. Moreover even for
the same combination of fluids it depends strongly on the depth ratio (cf. 
fig.\ref{ahnl2}). Nevertheless in most cases the values found are significantly
smaller than those characteristic for Marangoni convection in single layer
systems. In \cite{EBT} the transition from hexagons to squares in an
experiment with a single fluid layer were, e.g., reported to occur at 
$\epsilon\cong 4.2$ with the theoretical value resulting from a numerical
integration of the Navier-Stokes equation being even higher. 
For the two-layer setups 4 and 3 on the other hand studied in \cite{TMM} and
\cite{Anne} respectively 
$\epsilon_{hts}$ is so small that it is well conceivable to miss the hexagonal
pattern completely in the experiment and to observe squares as the first
pattern after the instability in accordance with experimental findings. 
For setup 3 one also notes that together with $\epsilon_{hts}$ also the
absolute value of $\epsilon_h$ characterizing the sub-critical stability region
of the hexagon planform gets very small such that hexagons exist only
in an extremely small window around criticality. Note 
also that our analysis is only concerned with {\it perfect} patterns hardly
occurring in the experiments. It seems well possible that squares are generated
by some inhomogeneous nucleation process even before $\epsilon_{hts}$ is
really reached. 

The remaining discrepancies between theoretical and experimental findings
might be due to the perturbative character of our derivation. In particular,
there is the possibility of so-called asymmetric squares in pattern forming
hydrodynamic systems \cite{BuCl} which, bifurcating {\it discontinuously} from the
quiescent state do not show up in a perturbative approach\footnote{We would
  like to thank F. Busse for pointing out this possibility to us.}. At the
moment it is not clear whether these patterns can be expected already at the
small values of the super-criticality parameter $\eps$ used in the
experiments. Since the flow pattern of asymmetric squares is rather different
from the one of conventional squares it might be possible to clarify
experimentally which form of squares has been observed.\\[.3cm]

{\bf Acknowledgment:} We have very much benefited from discussions with Anne 
Juel and Harry Swinney. A.E. would also like to thank F. Busse and W. Pesch
for interesting discussions and Jean Bragard, Wayne Tokaruk and Stephen Morris 
for very useful correspondence. Part of the work was done during a stay of
A.E. at the Center for Nonlinear Dynamics at the University of Texas at
Austin. He would like to thank all members of the Center for the kind
hospitality and the {\it Volkswagenstiftung} for financial support.  The work
of J.B.S. was supported by the NASA Office of Life and and Microgravity
Sciences Grant NAG3-1839.

\appendix
\section{Parameter Values}
In this appendix we have collected the values of the hydrodynamic parameters
used for the numerical calculations of the present paper. All five sets 
correspond to experimentally relevant combinations. Setup 1-3 have been
studied in \cite{Anne}. Setup 4 was investigated in \cite{TMM}
whereas setup 5 is from the classical work \cite{ZeRe}.\\[.5cm]

\begin{tabular}{|c||c|c||c|c||c|c||}\hline 
&\multicolumn{2}{c||}{setup 1}&\multicolumn{2}{c||}{setup 2}
&\multicolumn{2}{c||}{setup 3}\\[2mm]\hline
& lower fluid & upper fluid &lower fluid & upper fluid
     & lower fluid & upper fluid\\\hline
substance &HT135 &silicon oil &HT 70 &silicon oil& 
    acetonitrile & n-hexane \\\hline 
$h (mm)$& 2 & 1 &.92 &2.14 &1.775 & 2.725 \\\hline
$\rho (\frac{kg}{m^3})$&1730 &940 &1680 &920 &776 & 655 \\\hline
$\nu (\frac{m^2}{s})$&$1\cdot 10^{-6}$&$2\cdot 10^{-6}$
   &$5\cdot 10^{-7}$ &$5\cdot 10^{-6}$
   &$4.76\cdot 10^{-7}$&$4.58\cdot 10^{-7}$ \\\hline
$\kappa (\frac{J}{msK})$&.070 &.134 &.070 & .117 &.188 &.120\\\hline
$c_p(\frac{J}{kg K})$&962 &1498 &962 &1590 & 2230 & 2270 \\\hline
$\alpha (\frac{1}{K})$&$1.10\cdot 10^{-3}$ &$1.10\cdot 10^{-3}$
   &$1.10\cdot 10^{-3}$&$1.05\cdot 10^{-3}$
   &$1.41\cdot 10^{-3}$&$1.141\cdot 10^{-3}$\\\hline
$\frac{d\sigma}{dT} (N/m\,K)$ &\multicolumn{2}{c||}{$-5\cdot 10^{-5}$}
   &\multicolumn{2}{c||}{$-4.5\cdot 10^{-5}$}
   &\multicolumn{2}{c||}{$-1\cdot 10^{-4}$}\\\hline
\end{tabular}

\begin{tabular}{|c||c|c||c|c||}\hline 
&\multicolumn{2}{c||}{setup 4}&\multicolumn{2}{c||}{setup 5}\\[2mm]\hline
& lower fluid & upper fluid &lower fluid & upper fluid\\\hline
substance &FC75-FC104 & water & water & benzene\\\hline 
$h (mm)$& 1.28 & 2.78 &2.0 &1.0\\\hline
$\rho (\frac{kg}{m^3})$&1760 &998 &999 &885\\\hline
$\nu (\frac{m^2}{s})$&$8\cdot 10^{-7}$&$1\cdot 10^{-6}$
   &$1.14\cdot 10^{-6}$ &$7.77\cdot 10^{-7}$\\\hline
$\kappa (\frac{J}{msK})$&.063 &.586 &.59 &.1615\\\hline
$c_p(\frac{J}{kg K})$&1046 &4104 &4186 &1757\\\hline
$\alpha (\frac{1}{K})$&$1.4\cdot 10^{-3}$ &$2.06\cdot 10^{-4}$
   &$1.50\cdot 10^{-4}$&$1.06\cdot 10^{-3}$\\\hline
$\frac{d\sigma}{dT} (N/m\,K)$ &\multicolumn{2}{c||}{$-4.7\cdot 10^{-5}$}
   &\multicolumn{2}{c||}{$-5\cdot 10^{-5}$}\\\hline
\end{tabular}

\vspace{.3cm}
\noindent Table 2: Parameter values for the five different experimental
  setups studied in this paper. In addition for all setups $g=9.81 m/s^2$ was
  used. Note that the value of $d\sigma/dT$ is difficult to determine
  experimentally, the given values are therefore rough estimates or fitted
  from the linear analysis.

\section{Operator expansion and adjoint problem}
The decomposition (\ref{Lexp}) of the linear operator is not completely
straightforward for the Marangoni problem because the bifurcation parameter
$M$ not only occurs in the linear operator but also in the corresponding
boundary conditions. A transparent way to deal with the situation is to
include the boundary condition involving $M$ into the operator $L$ \cite{CL}, 
which is then written in the form 
\begin{equation}\label{defLaug}
L=\begin{pmatrix}
     \Delta^2  & c M \Delta_{\perp}& 0 & 0 & 0\\
         1     &      \Delta       & 0 & 0 &0 \\
         0     &       0  & \nu\Delta^2 &\alpha c M \Delta_{\perp}& 0\\ 
         0     &       0  &\frac{1}{\ka} & \chi\Delta& 0 \\
 \pa^2_z\arrowvert_{z=0} & 0 &-\eta\pa^2_z\arrowvert_{z=0}&0&-M\Delta_{\perp}\\
    \end{pmatrix}
\end{equation}
acting now on the correspondingly augmented state vector
\begin{equation}\label{defphiaug}
  \ph=\begin{pmatrix} w\\ \te \\ W \\ \Te \\ \te|_{z=0}\end{pmatrix}\quad.
\end{equation}
The operator is completed by the boundary conditions
\begin{align}
w&=\pa_z w=\te=0\quad\text{at}\quad z=-1\label{bcaug1}\\
w&=W=0\,,\,\pa_z w=\pa_z W\,,\,\te=\Te\,,\,\pa_z\te=\ka\pa_z\Te\,,\,
\quad\text{at}\quad z=0   \label{bcaug2}\\
W&=\pa_z W=\Te=0\quad\text{at}\quad z=a\label{bcaug3}
\end{align}
which differ from (\ref{basicbc1})-(\ref{basicbc3}) just by the omission of
the boundary condition involving $M$. We now easily find
\begin{equation}\label{defL_0}
L_0=\begin{pmatrix}
     \Delta^2  & c M_c \Delta_{\perp}& 0 & 0 & 0\\
         1     &      \Delta       & 0 & 0 &0 \\
         0     &       0  & \nu\Delta^2 &\alpha c M_c \Delta_{\perp}& 0\\ 
         0     &       0  &\frac{1}{\ka} & \chi\Delta& 0 \\
 \pa^2_z\arrowvert_{z=0} & 0 &-\eta\pa^2_z\arrowvert_{z=0}&0
    &-M_c\Delta_{\perp}\\
    \end{pmatrix}\quad,
\end{equation}
\begin{equation}\label{defL_1}
L_1=\begin{pmatrix}
         0     & c M_1 \Delta_{\perp}& 0 & 0 & 0\\
         0     &       0             & 0 & 0 & 0 \\
         0     &       0  &    0     &\alpha c M_1 \Delta_{\perp}& 0\\ 
         0     &       0  &    0     & 0 & 0 \\
         0     &       0  &    0     & 0 &-M_1\Delta_{\perp}\\
    \end{pmatrix}\quad,
\end{equation}
and 
\begin{equation}\label{defL_2}
L_2=\begin{pmatrix}
         0     & c M_2 \Delta_{\perp}& 0 & 0 & 0\\
         0     &       0             & 0 & 0 & 0 \\
         0     &       0  &    0     &\alpha c M_2 \Delta_{\perp}& 0\\ 
         0     &       0  &    0     & 0 & 0 \\
         0     &       0  &    0     & 0 &-M_2\Delta_{\perp}\\
    \end{pmatrix}\quad,
\end{equation}
where all three operators are completed by the boundary conditions
(\ref{bcaug1})-(\ref{bcaug3}). 

The adjoint operator is defined by 
$\langle \bar{\ph}|L \ph\rangle =\langle L^+\bar{\ph}|\ph\rangle$. Introducing 
the scalar product 
\begin{equation}\label{defscalprod}
\langle \bar{\ph}|\ph\rangle=\lim_{L\to\infty}\frac{1}{L^2}\int_{-L/2}^{L/2}
 dx\int_{-L/2}^{L/2}dy \left[\int_{-1}^0 dz (\bar{w}^* w + \bar{\te}^* \te) + 
   \int_{0}^a dz (\bar{W}^* W + \bar{\Te}^* \Te) 
    + \pa_z\bar{w}^*|_{z=0}\te|_{z=0} \right]
\end{equation}
we find after some partial integration that $L^+$ is given by 
\begin{equation}\label{defLadj}
  L^+=\begin{pmatrix}
     \Delta^2          &         1         & 0 & 0&0\\
 c M \Delta_{\perp}    &      \Delta       & 0 & 0&0\\
         0     &       0           & \nu\Delta^2 &\frac{1}{\kappa}&0\\ 
         0     &       0    &\alpha c M \Delta_{\perp}& \chi\Delta&0\\
       0 & -\pa_z|_{z=0}& 0 & \chi \pa_z|_{z=0} & -M\Delta_{\perp}\\
    \end{pmatrix}
\end{equation}
acting on the augmented vector
\begin{equation}
\bar{\ph}=\begin{pmatrix} 
  \bar{w}\\  \bar{\te} \\  \bar{W} \\  \bar{\Te} \\ \pa_z\bar{w}|_{z=0}
  \end{pmatrix}
\end{equation}
and completed by the boundary conditions
\begin{align}
\bar{w}&=\pa_z \bar{w}=\bar{\te}=0\quad\text{at}\quad z=-1\label{bcad1}\\
\bar{w}&=\bar{W}=0\,,\,\pa_z \bar{w}=\frac{1}{\rho}\pa_z\bar{W}\,,\,
\pa_z^2 \bar{w}=\nu\pa_z^2\bar{W}\,,\,
\bar{\te}=\frac{\chi}{\kappa}\bar{\Te}\,,\,
\quad\text{at}\quad z=0 \label{bcad2}\\ 
\bar{W}&=\pa_z \bar{W}=\bar{\Te}=0\quad\text{at}\quad z=a\quad.\label{bcad3}
\end{align}
It is, of course, possible to transform back the last line of $L^+$ into a
boundary condition and this is indeed advantageous to determine $\bar{\ph_0}$
explicitly, however for the use in the solvability conditions the above
augmented form is the most appropriate one.

\section{The $O(\eps^2)$-problem}
In this appendix we solve eq.(\ref{hier2}) for the case of a static
instability. From the term ${\cal  N}(\ph_0,\ph_0)$ 
and the structure (\ref{formph0}) of $\ph_0$ it is clear that the right hand
side of this equation will contain several terms with different exponential
factors of the form $\exp(i(\pm\bk_n\pm\bk_m)\br)$. Because of the linearity
of the equation we may solve it separately for all these term in the
inhomogeneity.  

Let us start with the so-called {\it non-resonant} terms in which the
angle $\phi$ between $\pm\bk_n$ and $\pm\bk_m$ is different from $2\pi/3$. It
is clear then from the $x$-$y$-integrals in (\ref{defscalprod}) that for these
terms $\langle \bar{\ph_0} | {\cal N}(\ph_0,\ph_0)\rangle=0$. In view of
(\ref{defL_1}) the solvability condition boils down to $M_1=0$ and hence
removes the $L_1 \ph_0$-term from the inhomogeneity of (\ref{hier2}). Using
the form (\ref{formph0}) of $\ph_0$ we therefore find as equations for $\ph_1$:
\begin{align*}\label{eq2nonres}
\Delta^2 w_1 + c M_c \Delta_{\perp}\te_1&=A_n A_m e^{i(\pm\bk_n\pm\bk_m)\br}\;
  \frac{2}{Pr}\left[(1+\cos(\phi))(w_0'''w_0+(1-2\cos(\phi))w_0'w_0'')-
  2 k_c^2\sin^2(\phi)\,w_0 w_0')\right]\\
w_1+\Delta \te_1&=A_n A_m e^{i(\pm\bk_n\pm\bk_m)\br}\;
  2\,(w_0 \te_0'-\cos(\phi)\,w_0'\te_0)\\
\nu\Delta^2 W_1 + \alpha c M \Delta_{\perp}\Te_1&=
       A_n A_m e^{i(\pm\bk_n\pm\bk_m)\br}\;
  \frac{2}{Pr}\left[(1+\cos(\phi))(W_0'''W_0+(1-2\cos(\phi))W_0'W_0'')-
  2 k_c^2\sin^2(\phi)\,W_0 W_0')\right]\\
\frac{1}{\ka}W_1+\chi\Delta\Te_1&=A_n A_m e^{i(\pm\bk_n\pm\bk_m)\br}\;
  2\,(W_0 \Te_0'-\cos(\phi)\,W_0'\Te_0)\qquad,
\end{align*}
where the prime denotes differentiation with respect to $z$. Since $M_1=0$ the 
boundary conditions completing this set of equations are given by
(\ref{basicbc1})-(\ref{basicbc3}) with $M=M_c$. 

The solution of these equations is of the form 
$\ph_1=A_n A_m\,\ph_1(z)\,\exp(i(\pm\bk_n \pm\bk_m)\br)$. 
We first determine a solution of the inhomogeneous equations using
the ans\"atze  
\begin{alignat}{2}
w_1^{inh}(z)&=\sum_{i,j=1}^6 w_{1ij} e^{(\la_i+\la_j)z}\qquad&\qquad
\te_1^{inh}(z)&=\sum_{i,j=1}^6 \te_{1ij} e^{(\la_i+\la_j)z}\\
W_1^{inh}(z)&=\sum_{i,j=7}^{12} w_{1ij} e^{(\la_i+\la_j)z}\qquad&\qquad
\Te_1^{inh}(z)&=\sum_{i,j=7}^{12} \te_{1ij} e^{(\la_i+\la_j)z}
\end{alignat}
which give rise to algebraic equations for the coefficients
$w_{1ij},\te_{1ij},W_{1ij}$ and $\Te_{1ij}$ in terms of $w_{0i}$ and
$\la_i$. This solution does not yet satisfy the boundary conditions. We
therefore add a proper solution of the homogeneous equation which is written
in the form 
\begin{alignat}{2}
w_1^{hom}(z)&=\sum_{i=1}^6 w_{1i}^{hom}\,e^{\tilde{\la}_i z}\qquad&\qquad
\te_1^{hom}(z)&=-\sum_{i=1}^6 \frac{w_{1i}^{hom}}
   {\tilde{\la}_i^2-2k_c^2(1+\cos(\phi))}\,e^{\tilde{\la}_i z}\\
W_1^{hom}(z)&=\sum_{i=7}^{12} w_{1i}^{hom}\,e^{\tilde{\la}_i z}\qquad&\qquad
\Te_1^{hom}(z)&=-\frac{1}{\kappa\chi} \sum_{i=7}^{12} 
 \frac{w_{1i}^{hom}}{\tilde{\la}_i^2-2 k_c^2 (1+\cos(\phi))}\,
 e^{\tilde{\la}_i z}
\end{alignat}
with $\tilde{\la}_i$ satisfying
\begin{equation}
[\tilde{\la}^2_i-2k_c^2\,(1+\cos(\phi))]^3=\left\{\begin{array}{lll}
   -2\,cM_c\,k_c^2 (1+\cos(\phi)) &\quad\text{for}\quad i&=1,\dots,6\\ & &\\
   -2\frac{\alpha}{\kappa\nu\chi}\,cM_c\,k_c^2\,(1+\cos(\phi)) 
      &\quad\text{for}\quad i&=7,\dots,12\end{array}\right.\quad.
\end{equation}
Note that $\tilde{\la}_i\neq\la_i$. Therefore the determinant of the matrix in 
the inhomogeneous set of linear equations for $w_{1i}^{hom}$ is different from
zero and the solution is unique. Note also that for $\phi=\pi$ the procedure
can be simplified since $w_1(z)=W_1(z)=0$. 

As for the {\it resonant} terms arising from the interaction of
modes with an angle $\phi=2\pi/3$ between their respective $\pm\bk$-vectors
let us focus on the one proportional to $\exp(i\bk_1\br)$. It is has one
contribution proportional to $A_1$ stemming from $-L_1\ph_0$ and another one
proportional to $A^*_2 A^*_3$ originating from ${\cal N}(\ph_0,\ph_0)$ in
(\ref{hier2}). Using $L_1$ as defined by (\ref{defL_1}) the resulting
equations are of the form 
\begin{align}
\Delta^2 w_1 + c M_c \Delta_{\perp}\te_1&=e^{i\bk_1\br}\,
 \frac{A^*_2 A^*_3}{Pr}(w_0'''w_0+2\,w_0'w_0''-
  3 k_c^2\,w_0 w_0')+A_1cM_1k_c^2\,\te_0\label{eq2res1}\\
w_1+\Delta \te_1&=e^{i\bk_1\br}\,A^*_2 A^*_3\,
  (2\,w_0 \te_0'+w_0'\te_0)\label{eq2res2}\\
\nu\Delta^2 W_1 + \alpha c M \Delta_{\perp}\Te_1&=e^{i\bk_1\br}\,
 \frac{A^*_2 A^*_3}{Pr}(W_0'''W_0+2\,W_0'W_0''-
  3 k_c^2\,W_0 W_0')+A_1\alpha cM_1k_c^2\,\Te_0\label{eq2res3}\\
\frac{1}{\ka}W_1+\chi\Delta\Te_1&=e^{i\bk_1\br}\,A^*_2 A^*_3\,
  (2\,W_0 \Te_0'+W_0'\Te_0)\quad.\label{eq2res4}
\end{align}
The boundary conditions are again given by (\ref{basicbc1})-(\ref{basicbc3})
except for the one containing the Marangoni number, which is modified to
(cf. \ref{defL_1})
\begin{equation}\label{modbc}
\pa^2_z w_1-\eta \pa^2_z W_1 -M_c \Delta_{\perp} \te_1=-A_1\,e^{i\bk_1\br}\,
     M_1\,k_c^2 \te_0\quad\text{at}\quad z=0\quad.
\end{equation}
Due to the resonant factor $e^{i\bk_1\br}$ the terms arising from
${\cal N}(\ph_0,\ph_0)$ are not automatically perpendicular to $\bar{\ph_0}$
and using (\ref{defscalprod}) the solvability condition acquires the
non-trivial form
\begin{align}\label{F2}
0=A^*_2 A^*_3\,&\left[\int_{-1}^0 dz \left(\frac{\bar{w}^*_0}{Pr}
 (w_0'''w_0+2\,w_0'w_0''-3 k_c^2\,w_0 w_0')+
  \bar{\te}^*_0(2\,w_0 \te_0'+w_0'\te_0)\right)\right.\\\nonumber
        + &\left.\int_0^a dz \left(\frac{\bar{W}_0^*}{Pr}
 (W_0'''W_0+2\,W_0'W_0''-3 k_c^2\,W_0 W_0')+
  \bar{\Te}^*_0(2\,W_0 \Te_0'+W_0'\Te_0)\right)\right]\\\nonumber
  +A_1\,cM_1\,k_c^2 &\left[\int_{-1}^0 dz \,\bar{w}^*_0\,\te_0
       +\alpha\int_0^a dz \,\bar{W}^*_0\,\Te_0 
       -\frac{1}{c}\,\pa_z\bar{w}^*_0|_{z=0}\,\te_0|_{z=0}\right]\quad.
\end{align}
We use this equation to replace the terms involving $M_1$ in
eqs.(\ref{eq2res1})-(\ref{eq2res4}) and in the boundary condition
(\ref{modbc}). The solutions to these equations can then be written in the
form $A^*_2 A^*_3\,\ph_1(z)\,e^{i\bk_1\br}$. Again we first determine a
particular solution of the inhomogeneous equations by using the ans\"atze:
\begin{alignat*}{2}
w_1^{inh}(z)&=\sum_{i,j=1}^6 w_{1ij} e^{(\la_i+\la_j)z}
              +\sum_{i=1}^6 w_{1i}\,z\,e^{\la_i z}\qquad&\qquad
\te_1^{inh}(z)&=\sum_{i,j=1}^6 \te_{1ij} e^{(\la_i+\la_j)z}
              +\sum_{i=1}^6 (\te_{1i}\,z+\tilde{\te}_{1i})\,e^{\la_i z}\\
W_1^{inh}(z)&=\sum_{i,j=7}^{12} w_{1ij} e^{(\la_i+\la_j)z}
              +\sum_{i=7}^{12} w_{1i}\,z\,e^{\la_i z}\qquad&\qquad
\Te_1^{inh}(z)&=\sum_{i,j=7}^{12} \te_{1ij} e^{(\la_i+\la_j)z}
              +\sum_{i=7}^{12} (\te_{1i}\,z+\tilde{\te}_{1i})\,e^{\la_i z}
              \quad.
\end{alignat*}
To satisfy the boundary conditions we add a solution of the homogeneous
equations which must be of the form (cf. (\ref{h3}),(\ref{h2}))
\begin{alignat}{2}
  w_1^{hom}(z)&=\sum_{i=1}^6 w_{1i}^{hom}\,e^{\la_i z}\qquad&\qquad 
\te_1^{hom}(z)&=-\sum_{i=1}^6 \frac{w_{1i}^{hom}}{\la_i^2-k_c^2}\,e^{\la_i z}\\
  W_1^{hom}(z)&=\sum_{i=7}^{12} w_{1i}^{hom}\,e^{\la_i z}\qquad&\qquad 
\Te_1^{hom}(z)&=-\frac{1}{\kappa\chi} \sum_{i=7}^{12} 
        \frac{w_{1i}^{hom}}{\la_i^2-k_c^2}\,e^{\la_i z}\quad.
\end{alignat}
The boundary conditions give rise to an {\it in}homogeneous system of
linear equations for the coefficients $w_{1i}^{hom}$ with the same singular
matrix ${\cal A}$ which appeared in the linear stability analysis. Due to the
solvability condition (\ref{F2}) however, the inhomogeneity of this set of
linear equations is perpendicular to the zero eigenvector of the adjoint
problem and therefore the system admits solutions. Their numerical
determination is most conveniently done by using the singular value
decomposition of the matrix ${\cal A}$ \cite{NR}. This method yields 
an approximate solution even if the solvability condition is not fulfilled
exactly, which will always be the case due to numerical errors. Moreover, the
so-called residual quantifying the deviation from the exactly solvable case 
gives another check of the numerical accuracy of the whole procedure.

Finally, the solution for $w_{1i}^{hom}$ obtained in this way is not unique
since one can always add a solution of the homogeneous equations. We will
enforce the additional constraint 
\begin{equation}\label{addconstr}
0=(\ph_0|\ph_1) := \lim_{L\to\infty}\frac{1}{L^2}\int_{-L/2}^{L/2}
 dx\int_{-L/2}^{L/2}dy \left[\int_{-1}^0 dz (w_0^* w_1 + \te_0^* \te_1) + 
   \int_{0}^a dz (W_0^* W_1 + \Te_0^*\Te_1)\right]
\end{equation}
to remove this ambiguity. The rationale behind this requirement is as
follows. Assume that we knew the exact solution $\ph$ of the full non-linear
problem. According to (\ref{pertans1}) and (\ref{formph0}) we want $A_n$ to be 
the amplitude of the contribution to $\ph$ proportional to $\exp(i\bk_n\br)$,
i.e. $(\exp(i\bk_n\br)\ph_0(z)|\ph)=\eps A_n$. Using the expansion
(\ref{pertans1}) for $\ph$ this results in $(\ph_0|\ph_l)=0$ for all 
$l\ge 1$. Note the use of different scalar products in (\ref{addconstr}) and
(\ref{defscalprod}).

This completes the solution of the $O(\eps^2)$ equations. The results are
specified by the various matrices 
$w_{1ij}, w_{1i}, w_{1i}^{hom}, \te_{1ij}, \te_{1i}, \tilde{\te}_{1i}$ and 
$\te_{1i}^{hom}$.

\end{document}